\documentclass[prl,twocolumn,superscriptaddress]{revtex4}
\usepackage{graphicx}
\usepackage{amssymb}
\usepackage{amsmath}
\usepackage{bm}
\usepackage{color}

\begin{document}

\title{Controlled ordering of topological charges in an exciton-polariton chain}

\author{T. Gao}
\affiliation{Nonlinear Physics Centre, Research School of Physics and Engineering, The Australian National University, Canberra ACT 2601, Australia}
\affiliation{Institute of Molecular plus, Tianjin University, 300072 Tianjin, China}

\author{O. A. Egorov}
\affiliation{Technische Physik, Wilhelm-Conrad-R\"ontgen-Research Center for Complex
Material Systems, Universit\"at W\"urzburg, Am Hubland, D-97074 W\"urzburg,
Germany}

\author{E. Estrecho}
\affiliation{Nonlinear Physics Centre, Research School of Physics and Engineering, The Australian National University, Canberra ACT 2601, Australia}
\affiliation{ARC Centre of Excellence in Future Low-Energy Electronics Technologies}

\author{K. Winkler}
\affiliation{Technische Physik, Wilhelm-Conrad-R\"ontgen-Research Center for Complex
Material Systems, Universit\"at W\"urzburg, Am Hubland, D-97074 W\"urzburg,
Germany}

\author{M. Kamp}
\affiliation{Technische Physik, Wilhelm-Conrad-R\"ontgen-Research Center for Complex
Material Systems, Universit\"at W\"urzburg, Am Hubland, D-97074 W\"urzburg,
Germany}

\author{C. Schneider}
\affiliation{Technische Physik, Wilhelm-Conrad-R\"ontgen-Research Center for Complex
Material Systems, Universit\"at W\"urzburg, Am Hubland, D-97074 W\"urzburg,
Germany}

\author{S. H\"ofling}
\affiliation{Technische Physik, Wilhelm-Conrad-R\"ontgen-Research Center for Complex
Material Systems, Universit\"at W\"urzburg, Am Hubland, D-97074 W\"urzburg,
Germany}
\affiliation{SUPA, School of Physics and Astronomy, University of St Andrews, St Andrews
KY16 9SS, United Kingdom}
\affiliation{ARC Centre of Excellence in Future Low-Energy Electronics Technologies}

\author{A. G. Truscott}
\affiliation{Laser Physics Centre, Research School of Physics and Engineering, The Australian National University, Canberra ACT 2601, Australia}

\author{E. A. Ostrovskaya}
\affiliation{Nonlinear Physics Centre, Research School of Physics and Engineering, The Australian National University, Canberra ACT 2601, Australia}
\affiliation{ARC Centre of Excellence in Future Low-Energy Electronics Technologies}


\begin{abstract}
We demonstrate, experimentally and theoretically, controlled loading of an exciton-polariton vortex chain into a 1D array of trapping potentials. Switching between two  types of vortex chains, with topological charges of the same or alternating sign, is realised by means of appropriate shaping of an incoherent pump beam that drives the system to the regime of bosonic condensation. In analogy to spin chains, these vortex sequences realise either a ``ferromagnetic" or an ``anti-ferromagnetic" order, whereby the role of spin is played by the orbital angular momentum. The ``ferromagnetic" ordering of vortices is associated with the formation of a persistent chiral current. Our results pave the way for  controlled creation of nontrivial distributions of orbital angular momentum and topological order in a periodic exciton-polariton system.
\end{abstract}
\maketitle

{\em Introduction.---} Microcavity exciton polaritons attract a great deal of interest as an accessible solid-state platform for fundamental studies of non-equilibrium macroscopic quantum systems \cite{DengREV10,CiutiREV13,YamamotoREV14,Book07}, as well as development of the polariton-based optoelectronics \cite{SanvittoREV16}. These quasiparticles arise due to hybridisation of electron-hole pairs (excitons) and photons in high-quality planar microcavities with embedded semiconductor quantum wells. Bosonic nature of exciton polaritons allows for spontaneous formation of macroscopic coherent states similar to Bose-Einstein condensates (BECs) \cite{Deng02,Kasprzak06,Snoke07,Nelson17}.

The spin degree of freedom \cite{Book07} and tunable interactions between multiply coupled polariton condensates have recently enabled the realisation of driven-dissipative bosonic spin lattices \cite{Ohadi16,Ohadi17}. In particular, a crossover from an antiferromagnetic state (with staggered spins) to a ferromagnetic state (with aligned spins) in such polaritonic lattices has been observed. It was shown that these periodic systems are analogous to the Ising spin model, the latter being very successful in describing a wide range of condensed matter phenomena. Furthermore, creation of periodic polariton arrays with controllable interactions between the nodes enabled demonstration of pseudo-spin lattices, which can be used as analogue simulators of $XY$ Hamiltonians \cite{Berloff17}.

In principle, the role of spin in the spin lattices and polariton simulators can be played by the {\em orbital angular momentum}, provided the latter is quantised, well-defined on a single lattice cell, and controllable. Being a quantum fluid, polariton condensate hosts a wide variety of quantum vortices \cite{Lagoudakis08,Roumpos11} with an integer {\em topological charge} defined by the phase winding around the vortex core. Polaritons optically injected by separated and independent incoherent pump spots can experience phase-locking, producing up to $10^2$ vortices and antivortices that extend over tens of microns across the sample and remain locked for a long time \cite{Tosi12}. However, the polariton vortex lattices demonstrated in experiments so far are characterised by spontaneously created ordering of topological charges, typically resulting in a zero net orbital angular momentum. Recent theoretical studies predict that the ability to control the ordering of topological charges in a periodic polariton lattice could offer information storage and processing capabilities \cite{Ma17}, as well as pave the way for realisation of topologically protected edge currents \cite{Liew17}.

\begin{figure}
\includegraphics[width=8.5 cm]{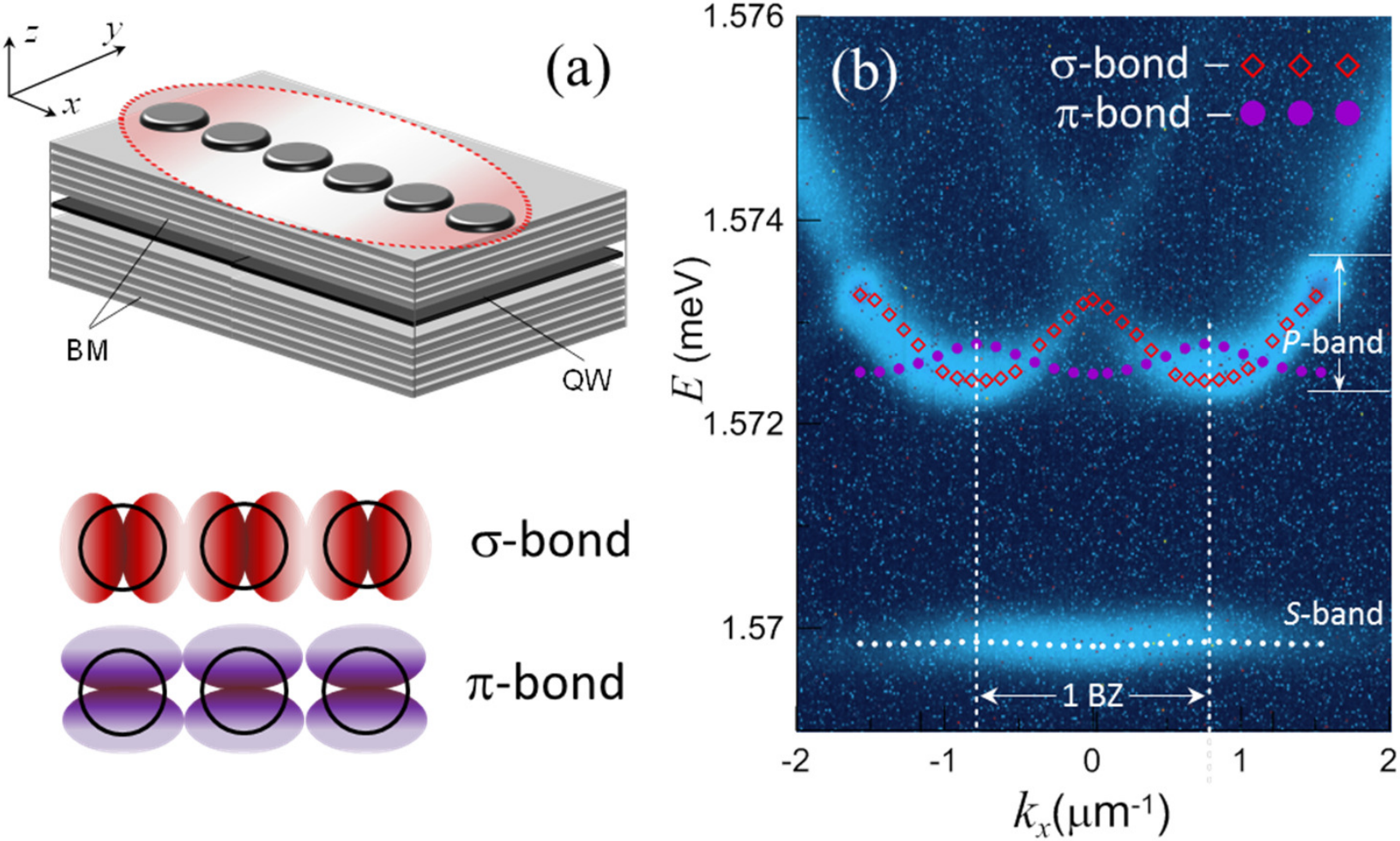}
\caption{(a) Schematics of a 1D mesa array microstructured in an AlAs/AlGaAs microcavity formed between two distributed Bragg mirrors (BM) with embedded GaAs quantum wells (QW) illuminated by an off-resonant optical pump (not to scale). (b) Dispersion measurement of the polariton emission from the mesa traps below condensation threshold. A $p$-energy band consist of two subbands and is formed by the hybridisation of the $p$-modes of the individual mesas. The inset in the right corner below: schematic representation of the $\sigma$-bond and $\pi$-bond hybridisation forming the $p$-band in panel (b). }
\label{Fig1_Sketch}
\end{figure}

In this work, we employ an exciton-polariton condensate in a 1D buried mesa array of polariton traps [see Fig.~\ref{Fig1_Sketch}(a)] \cite{Deveaud06,Winkler15, Winkler16,Schneider17} to observe the formation of exciton-polariton vortex chains. Due to the non-equilibrium character of condensation, controlled loading of polaritons into distinct energy bands can be realised by shaping the optical excitation beam \cite{Baboux16,Winkler16, Gao16}. Under the conditions of incoherent optical pumping, we observe vortex chains characterised by different distributions of the topological charge (orbital angular momentum) along the array. By shaping the pump beam, we demonstrate switching between the two vortex chains with topological charges of the same or opposite sign. Such topological states represent the vortex analogs of one-dimensional spin systems with ferromagnetic and anti-ferromagnetic configurations, whereby the role of spin is played by the orbital angular momentum at each lattice site.

{\em Experiment.---} The experiment is performed using 1D mesa arrays microstructured in an AlAs/AlGaAs microcavity with $12$ GaAs quantum wells, as described in \cite{Winkler15, Winkler16, Gao16}. Mesas of $d=2.0$ $\mu$m diameter are separated centre-to-centre by the distance of $a=4.0$~$\mu$m, with the effective polariton potential depth of $\sim 5.5$ meV for each mesa, as sketched schematically in Fig.~\ref{Fig1_Sketch}(a). The exciton-polariton condensate is formed spontaneously by pumping the microcavity with a {\em cw} laser, which injects free carriers well above the exciton energy.  The pump beam has a FWHM dimension of $20 \times 20$~$\mu$m, and illuminates approximately $5$ mesas in the array. The Rabi splitting in the sample is $\sim 11.5$ meV.

In the regime of low excitation powers, below condensation threshold, the dispersion (energy vs. in-plane momentum) of exciton polaritons reveals the band-gap structure imposed by the periodicity of the trapping potential in the lateral ($x$) direction, as described in \cite{Winkler16, Gao16}. Both the ground ($s$) and the excited ($p$) bands of the characteristic band-gap spectrum of extended Bloch states can be seen in Fig.~\ref{Fig1_Sketch}(b).
A large band gap is seen between the ground and the first excited bands, indicating proximity to the tight-binding regime. The most relevant physics arises due to the essentially 2D Bloch states forming via hybridisation of the higher-order bound states ($p$-modes) of the individual mesas, shown as a dotted lines in the Fig.~\ref{Fig1_Sketch} (b). Depending on a spatial orientation of the $p$-modes with respect to the array, they hybridise into two distinct Bloch subbands, namely, $\sigma$-bonding and $\pi$-bonding bands [see Fig.~\ref{Fig1_Sketch}(a), inset].

\begin{figure}
\includegraphics[width=8.5 cm]{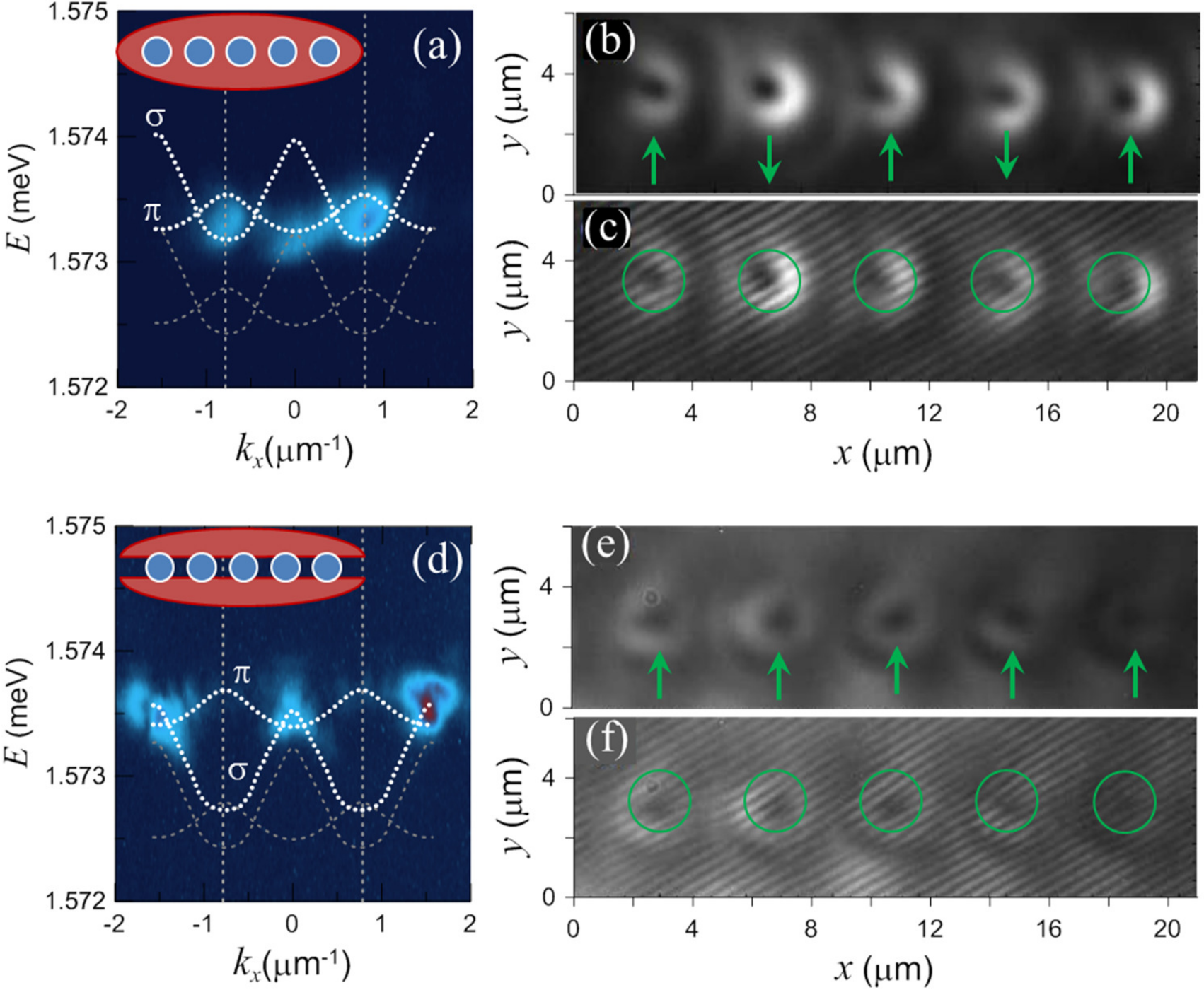}
\caption{Photoluminescence above condensation threshold. (a) Spectrum obtained under large incoherent optical beam (see sketch). Two distinct peaks at the edges of the BZ are visible. The signal in the centre of BZ was detected for non-vanishing $y$-components of the in-plane momentum ($k_y\neq0$). The dotted white line represent the blue-shifted dispersion subbands. (b) The intensity profile of the ``anti-ferromagnetic" vortex chain measured in real space and respective interferometry picture shown in (c). (d) Spectrum obtained under incoherent pumping outside the array shielded by a mask (see sketch). (e) The real-space intensity profile of the ``ferromagnetic" vortex chain and respective interferometry picture shown in (f). The on-site topological charges ($+1$ and $-1$) are represented by the up and down arrows, respectively.
}
\label{Fig2_Experiment}
\end{figure}

In the regime of strong excitation, the fast energy relaxation towards lower Bloch states is accompanied by the stimulated bosonic scattering into selected states with maximum gain \cite{Winkler16}. In the system under consideration, the second Bloch band, formed via bonding of the $p$-modes (dipoles) of individual mesas, becomes strongly populated as the power of the laser excitation grows [Fig.~\ref{Fig2_Experiment} (a),(d)]. As a rule, the condensation occurs in the vicinity of the high-symmetry points in the dispersion bands corresponding to zero group velocities \cite{Lai07,Cerda10,Tanese13}, either in the middle of the Brillouin Zone (BZ) or at its edges, depending on a particular spatial shape of the pumping spot. As shown below, the spatial distribution of the condensate in the form of the vortex chain can be interpreted as a phase-locked superposition of such 2D, higher-order Bloch states in the array. Therefore the formation of vortex states in our system is uniquely enabled by condensation of exciton polaritons into a non-ground Bloch state of the 1D array of 2D mesas. Moreover, depending on the pumping conditions, it is possible to populate Bloch states with different symmetries within the same band and, as a consequence, to switch between the vortex chains with ``staggered" [Fig.~\ref{Fig2_Experiment} (b),(c)] or ``aligned" [Fig.~\ref{Fig2_Experiment} (e),(f)] topological charges.

In the first configuration, the array of mesa traps is pumped by a large pump spot [Fig.~\ref{Fig2_Experiment}(a), inset]. At the higher pump powers, exciton polaritons undergo transition to bosonic condensation. Above-threshold dispersion indicates that the $\sigma$-bonding band is strongly occupied at the edges of the first BZ ($k_x=\pi/a$) and experiences blueshift due to interaction with incoherent excitonic reservoir injected by the pump [see the dotted white lines in Fig.~\ref{Fig2_Experiment}(a)]. An additional subset of occupied states, arising from the  $\pi$-bonding modes, is visible in the middle of the BZ ($k_x=0$) and is detectable only for the $k_y\neq0$ far-field components of the cavity photoluminescence. Real-space measurements [Figs.~\ref{Fig2_Experiment}(b),(c)] show a condensate distribution in the form of a chain of single-charge vortices with alternating sign of the topological charge ($-1,1,-1,1...$) or (using the spin-chain classification) ``anti-ferromagnetic" configuration. The distribution of the topological charges along the chain is confirmed by using the Michelson interferometry, whereby one arm of the interferometer undergoes $\times40$ magnification to create a defect-free reference beam [\ref{Fig2_Experiment}(c)].

In the second configuration, the array of mesa traps is pumped by a large pump spot, but the mesa array itself is shielded from the optical excitation by an optical mask [Fig.~\ref{Fig2_Experiment}(d), inset]. 
The dispersion measurement above condensation threshold shows that the exciton polaritons occupy the energy states in the middle of the first and the second BZs. In the real space and interferometry images, a vortex chain with the topological charges ($1,1,1,1...$) is observed [Figs.~\ref{Fig2_Experiment}(e),(f)], which corresponds to the ``ferromagnetic" ordering. Remarkably, this ``ferromagnetic" ordering signifies the formation of a {\em chiral state} of the exciton-polariton condensate in the 1D array due to non-zero net orbital angular momentum.

\begin{figure}
\includegraphics[width=8.5 cm]{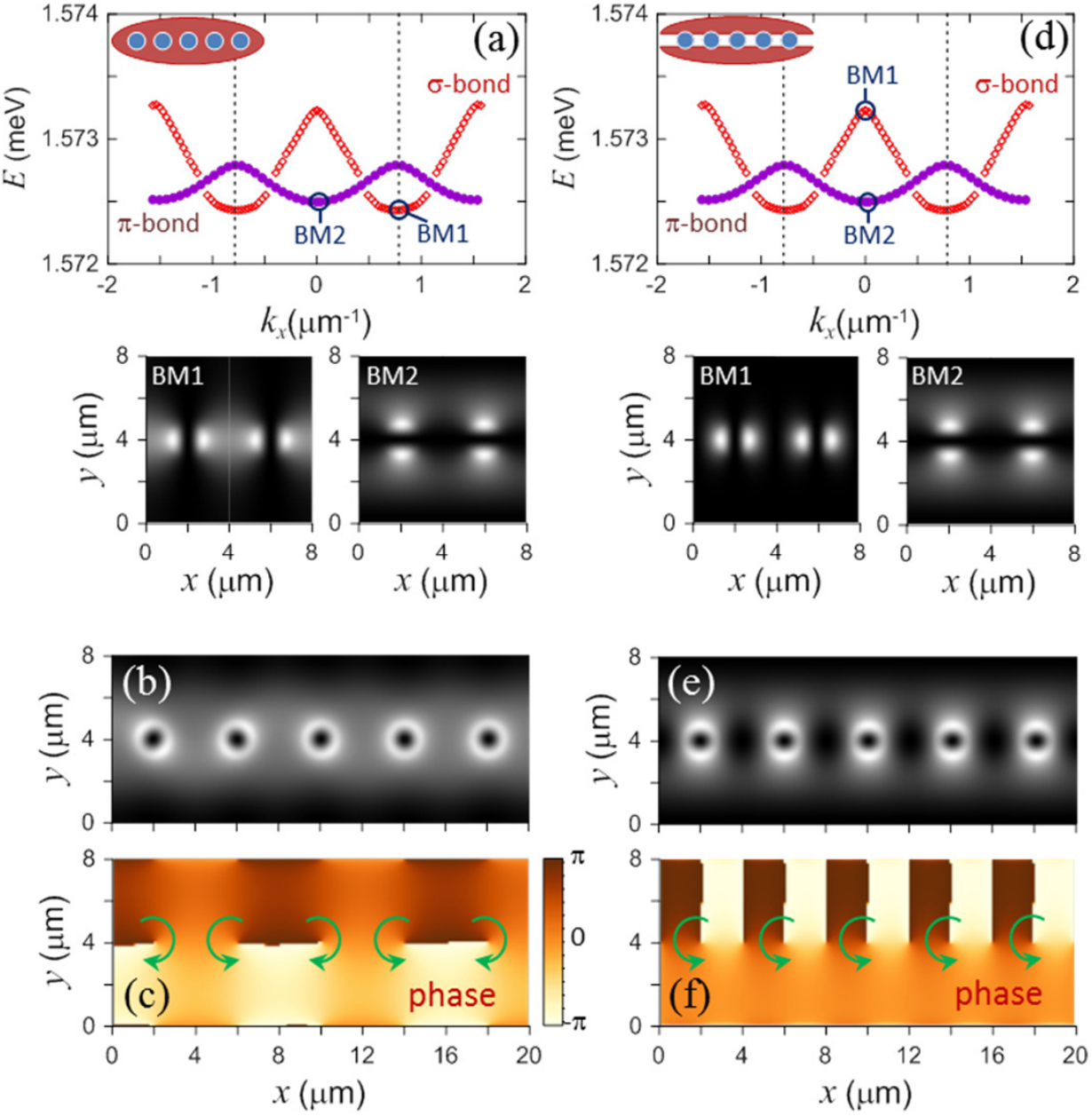}
\caption{Calculated Bloch modes dispersion of the $P$ band (a) and (d) and their liner superpositions (b),(c) and (e),(f). (a) $\sigma$- and $\pi$-subbands  of the chain. The circles show the Bloch states populated by a broad pump. The inset below shows the Bloch mode intensity profiles denoted ``BM1" and ``BM2" respectively.  (b) Intensity and (c) phase profiles of the linear superposition of the two Bloch modes marked by the circles in panel (a) with the fixed phase difference of $\pi/2$.
(d)The same as in panel (b) but with the mesa array shielded from the optical pump by a mask. (e) Intensity and (f) phase profiles of the linear superposition of the two Bloch modes marked by the circles in panel (d) with the fixed phase difference of $\pi/2$.
}
\label{Fig3_Theory}
\end{figure}

{\em Theory and Discussions.---} In what follows, we present a simple, intuitive theory of the observed effects. More precisely, we show that a phase-locked superposition of two Bloch states within the highly-populated second energy band of the array results in the formation of both the  ``ferromagnetic"  and ``anti-ferromagnetic" chains discussed above.

First, it is instructive to analyse the dispersion relation and the Bloch states of the array in the single-particle linear limit. Solving the standard eigenvalue problem in the photon-exciton basis with the trapping potential for photons \cite{Deveaud06}, we find the dispersion of the Bloch modes within both $\sigma$- and $\pi$-bonding subbands shown in Figs.~\ref{Fig3_Theory} (a),(d). Without loss of generality, it is convenient to represent the condensate in the polariton basis. Then we can assume that the condensed fraction of the exciton polaritons above threshold can be represented as a superposition of the two Bloch modes:
\begin{eqnarray}
\psi(x,y,t)= a(t) B_{\sigma} (x,y) e^{-i\omega_a(k_a) t +ik_a x}+\\ \nonumber
\quad \quad \quad \quad \quad  \quad \quad   b(t) B_{\pi}(x,y) e^{-i\omega_b(k_b) t +ik_b x}, \label{eq_ansatz}
\end{eqnarray}
where the order parameter can be characterised by the complex slowly-varying amplitudes $a(t)$ and $b(t)$ associated with the Bloch states $B_{\sigma}(x,y)$ and $B_{\pi}(x,y)$  in the $\sigma$- and $\pi$-subbands, respectively. Spatial profiles of the selected Bloch modes are shown in Figs.~\ref{Fig3_Theory} (a),(d). A spatial overlap of the Bloch modes with the pump beam determines their effective pumping rates \cite{Baboux16}.

A large pump spot overlaps strongly  with both of the subband components, and therefore both components of the condensate are equally occupied.  
Above condensation threshold the exciton polaritons condense into the respective minima of the subbands, namely at the edge of the BZ for $\sigma$-subband ($k_a=\pm\pi/a$) and in the middle of BZ for the $\pi$-subband ($k_b=0$). Note that the energy difference between these two states in the low-density (linear) limit is vanishingly small, i.e. $\omega_{\pi}(\pi/a)-\omega_{\sigma}(0)\approx0$. As discussed below, the nonlinear effects, such as polariton- and reservoir-polariton interactions, as well as local gain-saturation effects result in the blueshift and phase-locking between these two Bloch modes in the high-density regime. The respective superposition of the Bloch modes with equal amplitudes and a fixed phase of $\pi/2$ forms a sequence of vortices with the alternating charges, as shown in Figs. ~\ref{Fig3_Theory} (b),(c). Therefore, similar to the experiment, a phase-locked state of the two Bloch modes can form an ``anti-ferromagnetic"  chain of vortices.

The situation changes if the array itself is shielded from optical excitation. In this case, the $\pi$-subband has a larger overlap with the pump spot due to the spatial orientation of the on-site dipoles. This leads to an imbalance in the effective pumping rates of the two subbands [Fig.~\ref{Fig3_Theory}(d)].
As a result, the polaritons first condense in the $\pi$-subband in the middle of the BZ ($k_b=0$) and experience stronger reservoir-induced blue shift. This blue shift can compensate the initial linear frequency mismatch between the $\sigma$ and $\pi$ Bloch modes, given as $\omega_{\pi}(0)-\omega_{\sigma}(0)\neq0$ [see Fig.~\ref{Fig3_Theory} (d)]. Above the condensation threshold, a superposition of these two Bloch modes with the fixed phase ($\pi/2$) gives rise to the vortex chain with the same topological charges in the ``ferromagnetic" configuration [see Figs.~\ref{Fig3_Theory} (e),(f)].
\begin{figure}
\includegraphics[width=8.5 cm]{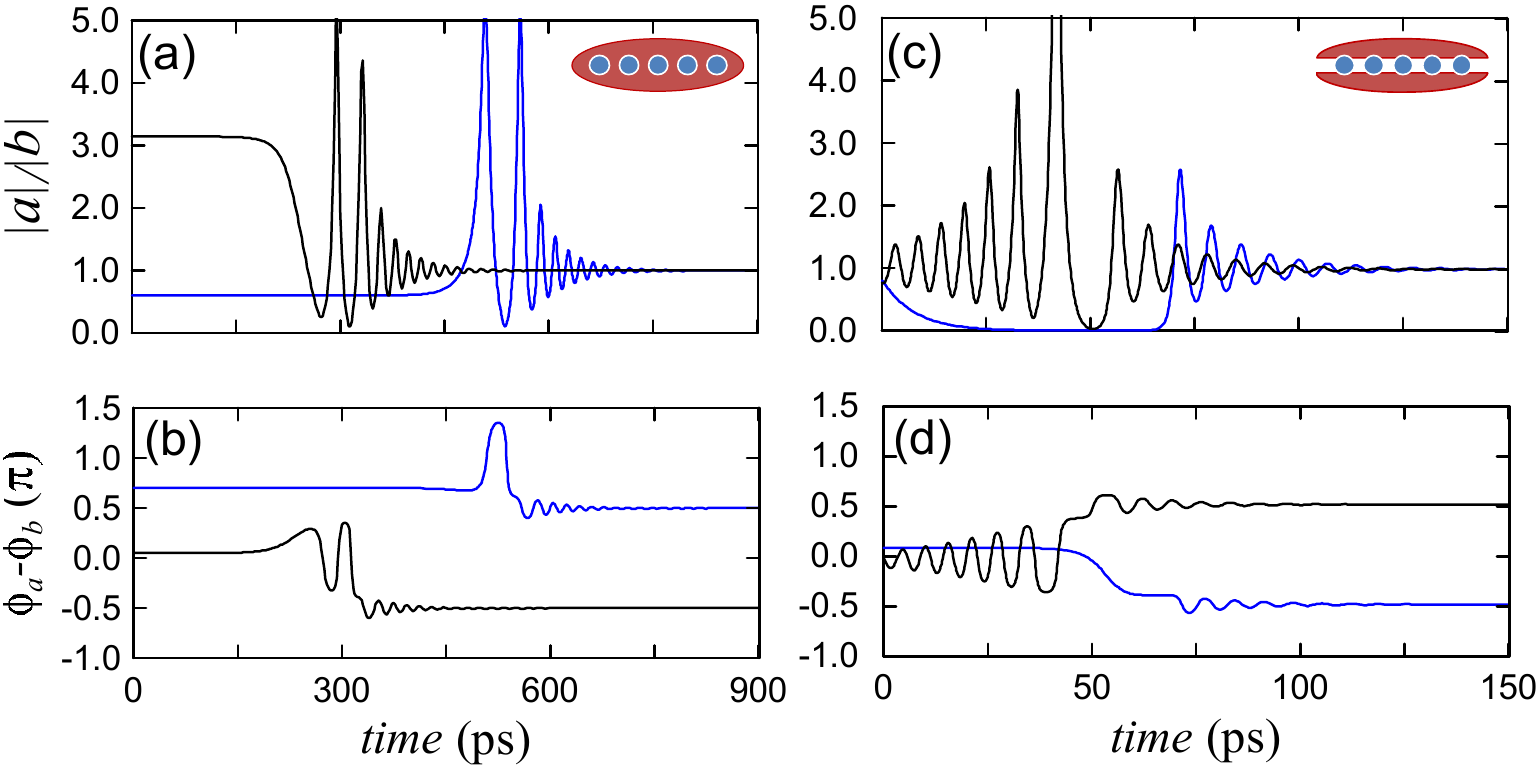}
\caption{Nonlinear dynamics of the Bloch mode amplitudes governed by the model~(\ref{eq_a},\ref{eq_b}). Panels (a) and (b) show the temporal dynamics of the amplitude ratio and the phase difference between the two Bloch modes, respectively, for the case of balanced pumping $P_{a}=P_{b}=35$ meV/ $\mu$m$^2$  and zero frequency detuning $\Delta_{ab}=\Delta_b-\Delta_a=0$. Panels (c) and (d) show the temporal dynamics of the amplitude ratio and the phase difference between the two Bloch-modes, respectively, for the case of imbalanced pumping $P_{a}<P_{b}$  ($P_{a}=32$ meV/$\mu$m$^2$, $P_{b}=45$ meV/$\mu$m$^2$)  and non-zero frequency detuning $\Delta_{ab}=0.63$ meV. For other parameters of the model see the Supplemental Materials ~\cite{SM}.
 }
\label{Fig4_Model}
\end{figure}

In order to understand the phase-locking mechanism qualitatively discussed above, it is necessary to consider the nonlinear effects. The full dynamics of exciton-polariton condensation in the one-dimensional mesa array for moderate pump powers above threshold can be reliably reproduced by the two-dimensional mean-field dynamical model taking into account energy relaxation due to quantum and thermal fluctuations in the system \cite{Wouters07, Wouters08}. It consists of the open-dissipative Gross-Pitaevskii equation for the condensate wave-function incorporating stochastic fluctuations and coupled to the rate equation for the excitonic reservoir created by the off-resonant cw pump \cite{Wouters08,Winkler16}. For the sake of simplicity, it is convenient to assume that the incoherent reservoir rapidly reaches the stationary density. Under this assumption, the reservoir density can be expressed explicitly and the model becomes closely related to that derived in ~\cite{Keeling08}. Then, by using the ansatz in the form~\eqref{eq_ansatz}, the standard model \cite{Wouters07} for the 
incoherently pumped polaritons with a relevant normalization can be written as (for the detailed derivation of the model see the Supplemental Materials ~\cite{SM}):
\begin{eqnarray} \label{eq_a}
i\frac{{d{a}}}{{dt}} = -\Delta _a {a} + \Gamma{a}+\left( {{\xi _a}{{\left| {{a}} \right|}^2} + 2{\xi _{ab}}{{\left| {{b}} \right|}^2}} \right){a} +{\tilde\xi _{ab}}b^2a^*,\nonumber
\end{eqnarray}
\begin{eqnarray} \label{eq_b}
i\frac{{d{b}}}{{dt}} = -\Delta _b {b} + \Gamma_b{b}+\left( {{\xi _b}{{\left| {{b}} \right|}^2} + 2{\xi _{ab}}{{\left| {{a}} \right|}^2}} \right){b} +{\tilde\xi _{ba}}a^2b^*,\nonumber
\end{eqnarray}
where $\Gamma_{a,b}=\left( iR/2 +g_r/\hbar\right)P_b/\gamma_r - i{\gamma _c}/2$, $\gamma_c$ is the loss rate for the condensed polaritons, $g_r$ accounts for the strength of the blue shift of the coherent polaritons due to their interaction with the incoherent reservoir, $\gamma_r$ is the loss rate of the incoherent reservoir particles, $R$ is the stimulated scattering rate into the condensed state, $P_a$, $P_{b}$ are the effective pump rates for the respective Bloch modes, and $\Delta_{a,b}=\omega_{a,b}-\omega_{\sigma,\pi}$. The complex coefficients $\xi _a$, $\xi _b$, $\xi _{ab}$, $\tilde{\xi}_{ab}$, $\tilde{\xi}_{ba}$ account for both polariton-polariton and reservoir-polariton interactions, as well as the effect of gain saturation due to reservoir depletion.  

First, we consider the ``anti-ferromagnetic" configuration which is associated with a balanced excitation of both Bloch modes ($P_{a}=P_{b}$). Extensive numerical simulations of the model~\eqref{eq_a},\eqref{eq_b} with random initial conditions proves the existence of a stable steady-state solution with equal amplitudes $|a|=|b|$ and a fixed phase difference between Bloch modes  $\phi_a-\phi_b=\pi/2 +\pi n$, where $n$ is integer number [as shown in Figs.~\ref{Fig4_Model}(a),(b)]. This confirms the existence of a phase-locked superposition of Bloch modes resulting in the ``anti-ferromagnetic" vortex chain.

In the case of the ``ferromagnetic" configuration, as discussed above, there is an imbalance in the effective pumping of the two subbands, i.e. $P_{a}<P_{b}$. Numerical simulations for the Bloch mode with the non-zero frequency detuning $\Delta _{ab}=\Delta_{b}(0)-\Delta_{b}(0)\neq0$ and imbalanced pumping show that there exist stable states with almost equal amplitudes and locked phase $\phi_a-\phi_b\approx  \pi /2+\pi n$ [as seen in Figs.~\ref{Fig4_Model}(c),(d)].

{\em Conclusion.---} To summarise, we experimentally demonstrated vortex chains with both ``ferromagnetic" and ``anti-ferromagnetic" distributions of topological charges in an exciton-polariton condensate loaded into a 1D array of mesa traps, and supported our observations by a coupled-mode theory for the Bloch states.
This research opens an avenue for using the current advanced nanofabrication techniques to create and manipulate stable exciton-polariton vortex chains in microstructured potentials. Furthermore, controlled loading of the exciton-polariton condensate into a topologically ordered state offers interesting possibilities for future realisation of topologically protected exciton-polariton currents \cite{Liew17}.


\end{document}